\documentclass[aip,twocolumn,10pt]{revtex4}%
\usepackage{amsmath}
\usepackage{amsfonts}
\usepackage{amssymb}
\usepackage{graphics}
\usepackage{graphicx}
\providecommand{\U}[1]{\protect\rule{.1in}{.1in}}

\begin{document}
\title{Time-multiplexed amplification in a hybrid-less and coil-less Josephson parametric converter}
\author{Baleegh Abdo}
\author{Jose M. Chavez-Garcia}
\author{Markus Brink}
\author{George Keefe}
\author{Jerry M. Chow}
\affiliation{IBM T. J. Watson Research Center, Yorktown Heights, New York 10598, USA.}
\date{\today }

\begin{abstract}
Josephson parametric converters (JPCs) are superconducting devices capable of performing nondegenerate, three-wave mixing in the microwave domain without losses. One drawback limiting their use in scalable quantum architectures is the large footprint of the auxiliary circuit needed for their operation, in particular, the use of off-chip, bulky, broadband hybrids and magnetic coils. Here, we realize a JPC which eliminates the need for these bulky components. The pump drive and flux bias are applied in the new device through an on-chip, lossless, three-port power divider and on-chip flux line, respectively. We show that the new design considerably simplifies the circuit and reduces the footprint of the device while maintaining a comparable performance to state-of-the-art JPCs. Furthermore, we exploit the tunable bandwidth property of the JPC and the added capability of applying alternating currents to the flux line in order to switch the resonance frequencies of the device, hence demonstrating time-multiplexed amplification of microwave tones that are separated by more than the dynamical bandwidth of the amplifier. Such a measurement technique can potentially serve to perform time-multiplexed, high-fidelity readout of superconducting qubits. 
\end{abstract}

\maketitle

\newpage
The Josephson parametric converter (JPC) is a valuable resource in the measurement and processing of quantum information carried by microwave signals \cite{JPCnaturePhys,JPCnature,JPCreview}. It is used to perform high-fidelity, quantum nondemolition measurement of superconducting qubits \cite{JPCparity}, track their quantum trajectories in realtime \cite{QubitJPC,FluxoniumJumps}, enable feedback \cite{FeedbackJPC}, transduce quantum information via noiseless frequency conversion \cite{Conv,QuantumNode}, and generate two-mode squeezed states of the microwave field \cite{TwoModeSqueezing}. JPCs also have the potential of serving as remote entanglers \cite{RemoteEntangler} in distributed quantum networks. 

However, despite the many useful applications of the JPC, state-of-the-art JPCs \cite{Jamp,JPCreview,CompactMixer} suffer from three main limitations that hinder their use in scalable quantum architectures, i.e., narrow dynamical bandwidth on the order of $10$ $\operatorname{MHz}$ at $20$ dB of gain, low saturation input power on the order of a few microwave photons per inverse dynamical bandwidth, both of which do not enable the readout of more than one qubit per JPC, and lastly the large footprint of the auxiliary drive and bias circuit needed for the operation of the device. 

\begin{figure}
	[tb]
	\begin{center}
		\includegraphics[
		width=\columnwidth 
		]%
		{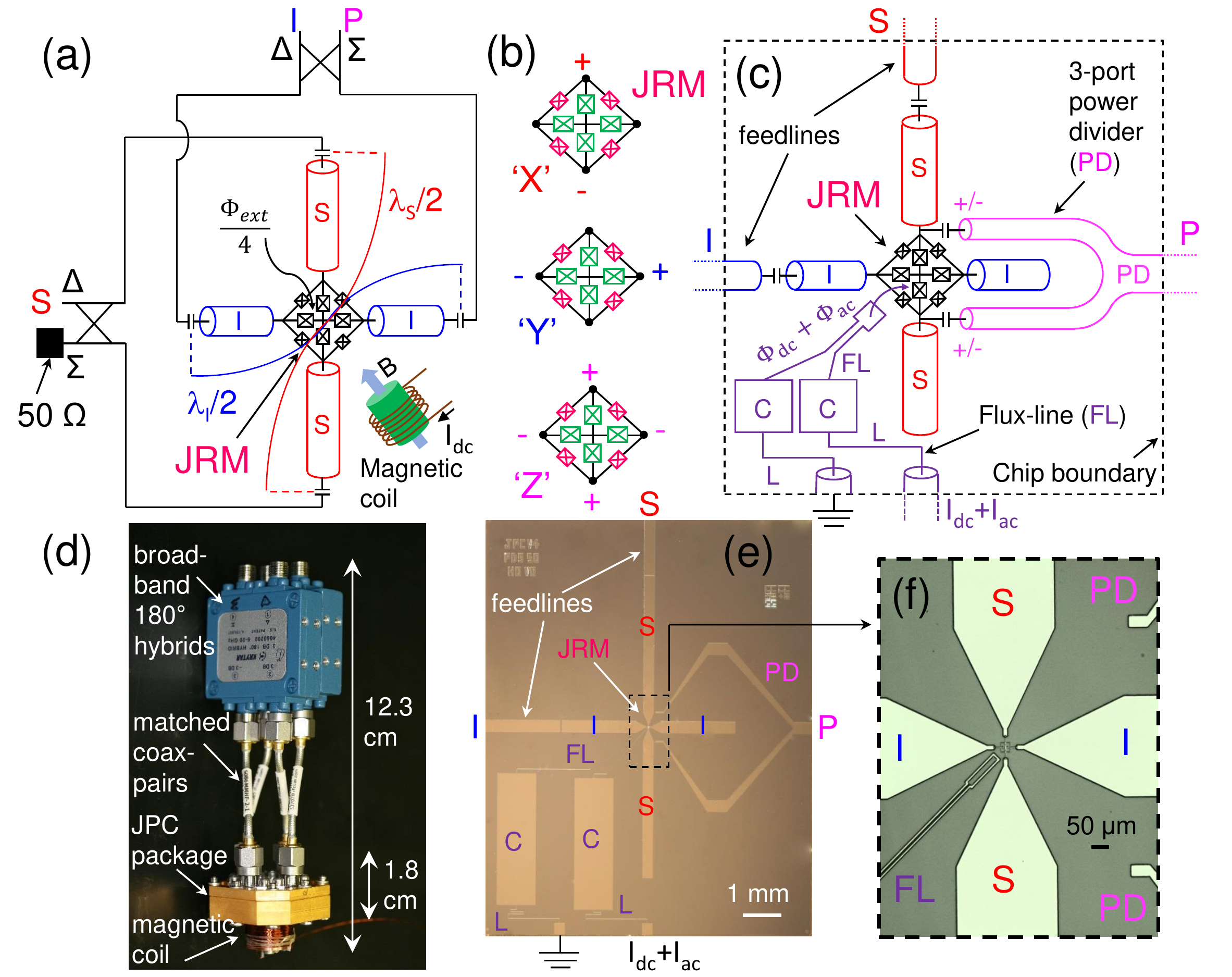}
		\caption{(color online). (a) State-of-the-art JPC circuit. The JPC consists of two microwave resonators with different fundamental resonance modes denoted signal (S) and idler (I), which intersect in the middle at a Josephson ring modulator (JRM). The device circuit also includes two off-chip, broadband $180$ degree hybrids and an external magnetic coil. (b) An illustration of the excitation pattern for the three eigenmodes of the JRM. (c) A block diagram of the new JPC circuit which eliminates the need for off-chip, bulky hybrids and coils. The P drive is fed through an on-chip, three-port power divider (PD) which is capacitively coupled to the JRM. The new design also includes an on-chip flux line (FL) which enables biasing the JRM using dc and ac currents. The FL incorporates a basic low-pass filter on each side which consists of a series inductance (L) in the form of a long and narrow superconducting wire and a shunting capacitance (C) to ground in the form of a large superconducting pad in order to reduce power leakage from the S and I resonators to the FL port. (d) A photo of the JPC device featuring the JPC package, the bulky hybrids, and the external magnetic coil. (e) A photo of the new JPC chip implemented on a $330$ $\mu\operatorname{m}$ thick silicon substrate. All conducting elements of the chip are made of niobium, except the JRM made of aluminum and the ground plane on the back of the substrate made of silver. (f) A blown-up photograph of the center region.            
		}
		\label{Device}
	\end{center}
\end{figure}

Here, we report on a new design and device which significantly simplifies the circuit and reduces the footprint of the JPC without degrading its performance. The new design is expected to not only  promote scalability, but also to make it possible to realize quantum-limited directional amplifiers and noiseless circulators (as have been featured in recent works \cite{NoiselessCirc,DircJPC,JDA,ReconfJJCircAmpl}). Furthermore, we  demonstrate using the new device, a time-multiplexed amplification technique, which can ,in certain scenarios, extend the utility of the amplifier bandwidth.    

To recognize the important aspects of the new JPC and its additional functionality, it is crucial to first briefly review the standard JPC physics and circuit. The JPC is a quantum-limited, nondegenerate, three-wave mixing device with both its signal and idler modes spatially and spectrally separated \cite{JPCreview}. In Fig. \ref{Device} (a) we show a circuit diagram of a state-of-the-art JPC. The main component of the JPC is the Josephson ring modulator (JRM) which functions as a nonlinear, dispersive mixing element. The JRM consists of four Josephson junctions (JJs) arranged in a Wheatstone bridge configuration (the outer loop). The JRM supports three microwave eigenmodes \cite{JPCnaturePhys} whose excitation patterns are illustrated in Fig. \ref{Device} (b), namely, two differential modes denoted `X' and `Y' and one common denoted `Z'. The dc circulating current in the outer loop induced by an external magnetic flux threading the JRM facilitates the mixing operation between the three modes \cite{JPCnaturePhys}. The other four large JJs inside the JRM serve as a linear shunt inductance for the outer JJs. The role of this inductive shunting is to lift the hysteretic response of the JRM versus flux and make the resonance frequencies of the JPC tunable \cite{Roch}. In order to couple the JRM to microwave signals and enhance the mixing interaction, the JRM is embedded at the center of two orthogonal, half-wavelength microstrip transmission line resonators denoted signal (S) and idler (I) which are capacitively coupled to external feedlines that carry microwave signals into and out of the JPC and whose fundamental resonance modes couple strongly to the JRM modes `X' and `Y' respectively. 

When operating the JPC as a quantum-limited parametric amplifier for S and I tones at frequencies $f_S$ and $f_I$ which lie within the dynamical bandwidth $B/2\pi$ of the device, a non-resonant common drive denoted pump (P) (which couples to excitation `Z') is applied to the device at frequency $f_P=f_S+f_I$ \cite{JPCnature,Jamp}. In order to address the S and I modes and apply the pump drive, both S and I tones are fed through the delta ports of two off-chip, broadband 180 degree hybrids, while the P is fed through the sigma port of one of the hybrids (the other sigma port is terminated by a $50$ Ohm cold load). As seen in the device photo shown in Fig. \ref{Device} (d), these hybrids (shown are Krytar Model 4060200) can be large in size. It is worth noting that at least one of the hybrids needs to be broadband in order to support either $f_S$ and $f_P$ or $f_I$ and $f_P$ which are in general separated by several gigahertz. Another disadvantage of this drive scheme is the insertion loss of the hybrids and the short matched coax cables connecting them to the JPC, which can be in the range 0.5-1.5 dB.         

\begin{figure}
	[tb]
	\begin{center}
		\includegraphics[
		width=0.7\columnwidth 
		]%
		{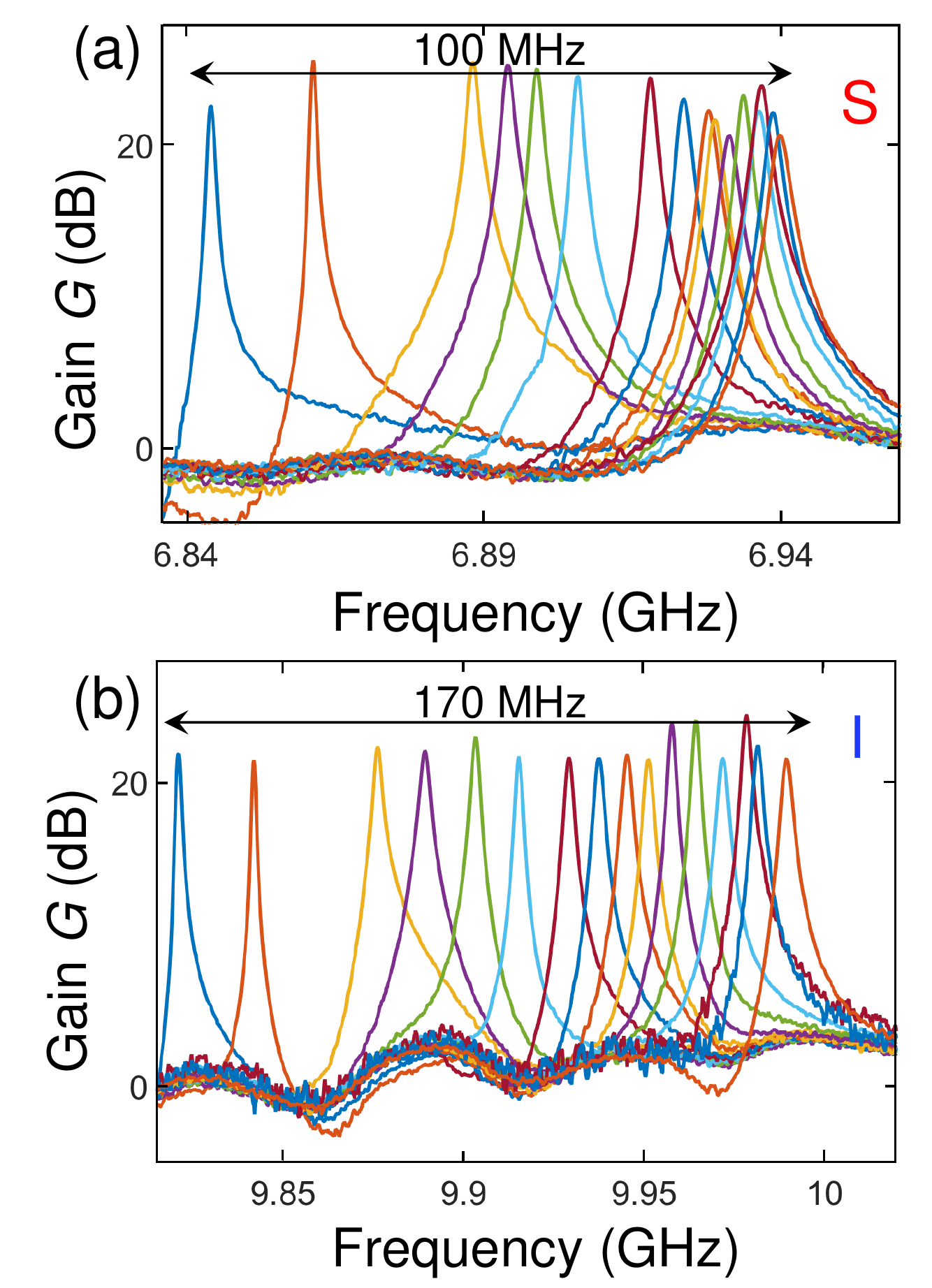}
		\caption{(color online). Tunable bandwidth measurement. The device gain curves are measured for various flux bias points at the S (a) and I (b) ports. The flux bias is varied by applying a dc current to the FL. The pump frequency and power are adjusted to yield a gain in excess of $20$ dB at each flux bias point.
		}
		\label{Gain}
	\end{center}
\end{figure}

\begin{figure*}
	[t!]
	\begin{center}
		\includegraphics[
		width=1.6\columnwidth 
		]%
		{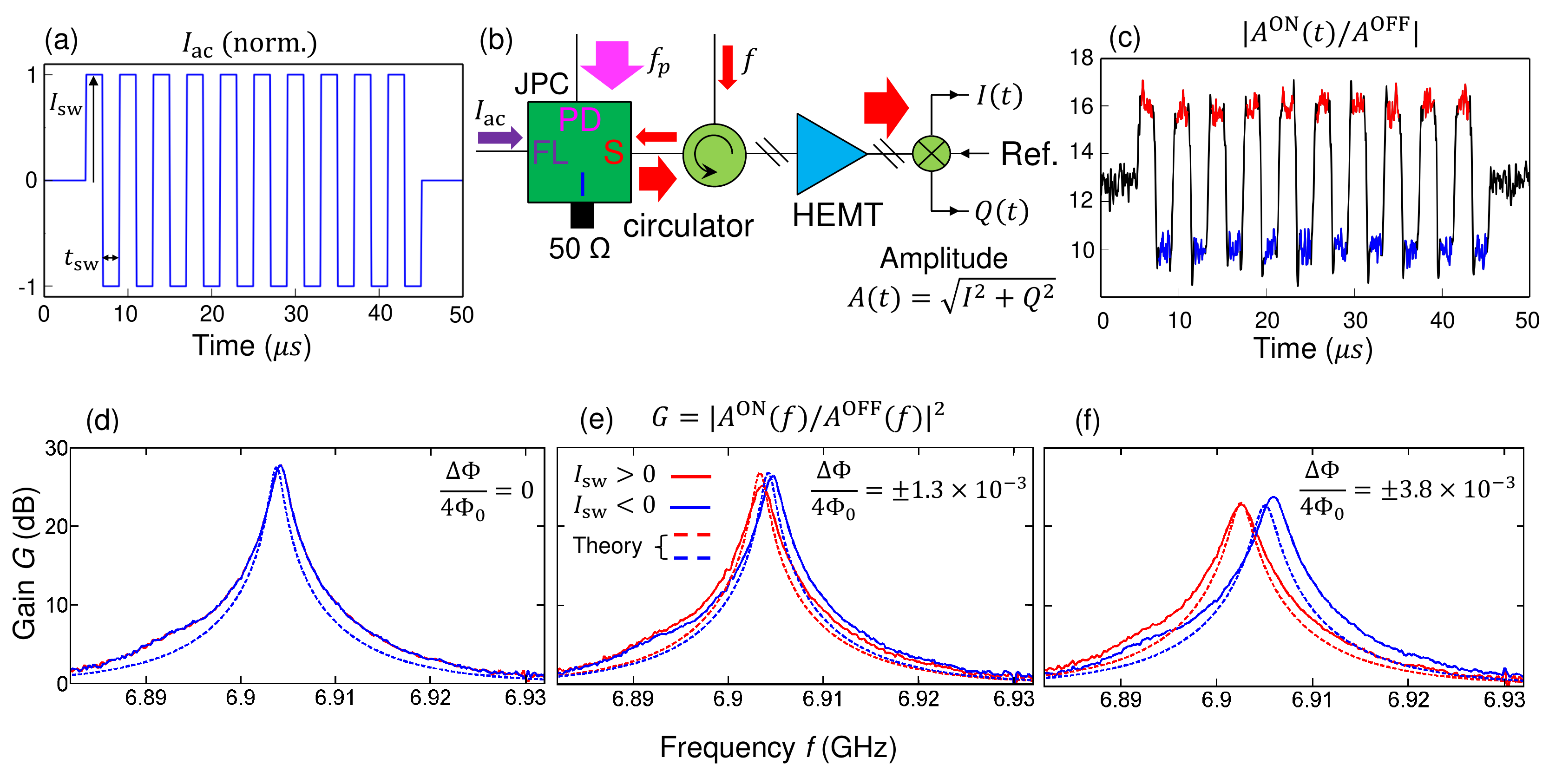}
		\caption{(color online). Time-multiplexed amplification measurement. (a) The waveform of the current square pulses applied to the FL with duration $t_{\mathrm{sw}}=2$ $\mu\operatorname{s}$ and varying amplitude $I_{\mathrm{sw}}$. (b) Block diagram of the measurement setup featuring the main components. The device working point is set by applying a flux bias to the JRM using an external magnetic coil and a pump drive at frequency $f_P=16.8195$ $\operatorname{GHz}$, which give a gain of $28$ dB at $f_S=6.904$ $\operatorname{GHz}$. In order to measure the JPC response to the current square pulses, a weak probe at frequency $f$ is applied to the signal port of the JPC. The frequency $f$ is swept within a span of $50$ $\operatorname{MHz}$ around $f_S$. The reflected amplified signal at $f$ is routed to the output using a circulator and amplified using a HEMT and a room-temperature amplifier before it is mixed down, filtered, and sampled using a digitizer. (c) A representative, normalized amplitude response of the output probe signal near resonance versus time, where the normalized amplitude corresponds to the ratio of the amplitude measured with and without pump drive and current pulses, denoted $A^{\mathrm{ON}}(t)$ and $A^{\mathrm{OFF}}$ respectively. In order to construct the red and blue gain curves versus frequency shown in panels (d)-(f), which correspond to the positive ($I_{\mathrm{sw}}>0$) and negative ($I_{\mathrm{sw}}<0$) current pulses, the respective steady state regions of the normalized amplitude response (colored red and blue in panel (c)) are averaged and squared. Panels (d)-(f) exhibit the constructed gain curves (solid lines) versus frequency for different $I_{\mathrm{sw}}$ which induce different flux variation in the JRM $\Delta\Phi$. The dashed lines represent calculated gain curves based on the device parameters and working point (see supplementary material). The values for $\Delta\Phi/4\Phi_0$, listed in the corner of plots (d)-(f), represent the normalized change in the flux threading one quarter of the JRM.      
		}
		\label{FastFreqSwitching}
	\end{center}
\end{figure*}

Our new JPC design aims to reduce the overall footprint and minimize the loss of off-chip components as schematically shown in Fig. \ref{Device} (c). In this scheme, we eliminate the broadband hybrids and short coaxes by implementing the S and I resonators in the form of open-ended half-wavelength transmission lines which are each capacitively coupled to a single feedline \cite{Roch}. As for feeding the pump drive, we realize an on-chip, lossless, three-port power divider (PD) which is a variation on the T-junction power divider \cite{Pozar}, that couples capacitively and symmetrically to two opposite nodes of the JRM (see supplementary material for another method of injecting the pump drive which yields similar results). In our case, the two arms of the PD couple to the top and bottom nodes as seen in Fig. \ref{Device} (e), (f). In designing the PD, three main considerations are taken into account, 1) keeping the two arms of the PD symmetrical, so that the injected pump is split evenly and in-phase between the two arms and, consequently inducing the desired `Z' excitation of the JRM, 2) capacitively coupling the two arms of the PD as close as possible to the JRM, located at an rf-voltage node of the S and I modes, in order to reduce power leakage through the PD, and to the same end 3) keeping the two arms of the PD as far as possible from either resonator.    

In addition, we implement an on-chip flux line (FL) [see Fig. \ref{Device}(c), (e), (f)], with a mutual inductance to the JRM of about $0.6$ p$\operatorname{H}$. The FL allows us to flux bias the JRM without using external magnetic coils attached to the JPC package (see Fig. \ref{Device}(d)). In addition to reducing the footprint of the JPC bias circuit, the FL has the advantage of supporting ac-currents/voltages and not only dc. Such capability can enable fast control of the amplifier gain and frequency as we show below. We have also incorporated into each side of the FL a basic low-pass, filtering element \cite{Pozar} with a designed cutoff frequency around $3$ $\operatorname{GHz}$ aimed at reducing power leakage from the S and I resonators through the FL (see Fig. \ref{Device} (e)).     

In Fig. \ref{Gain} (a) and (b), we exhibit a tunable bandwidth measurement of the new device in which we measure the power gain $G$ on the S and I ports respectively, for various flux bias points. In this measurement, the pump is fed through the PD while the flux bias points are set via a dc current applied to the FL. Additional figures of merit of the device such as, gain, bandwidth, added noise, and maximum input power are included in the supplementary material and exhibit comparable performance to state-of-the-art JPCs. 
      
In the measurements of Fig. \ref{FastFreqSwitching}, we take advantage of the tunable bandwidth property of the JPC and the added capability of applying time varying currents to the FL in order to demonstrate time-multiplexed amplification between two adjacent signal frequencies which are separated by more than the dynamical bandwidth of the JPC. Such fast frequency switching can be used for example to time-multiplex the readout of two qubit-cavity systems, whose readout frequencies are adjacent but do not lie within the dynamical bandwidth of the amplifier.   

To demonstrate time-multiplexed amplification, we set the JPC working point to $28$ dB of gain at $6.904$ $\operatorname{GHz}$ and measure the JPC response in the time domain while applying current square pulses of duration $t_{\mathrm{sw}}=2$ $\mu\operatorname{s}$ to the FL with alternating sign (Fig. \ref{FastFreqSwitching} (a)). In this measurement, the relation $t_{\mathrm{sw}}=2$ $\mu\operatorname{s}$ $\gg1/B=80$ n$\operatorname{s}$ ensures that the JPC reaches steady state during each pulse. In Fig. \ref{FastFreqSwitching} (b), we show the main components of the setup used in measuring the device response for varying input probe frequency $f$. By separately averaging the steady-state, time-domain output signal illustrated in Fig. \ref{FastFreqSwitching} (c) for the positive and negative current pulses over $10^{4}$ runs, we obtain two amplitudes for each $f$ drawn using red and blue lines respectively. In Fig. \ref{FastFreqSwitching} (d)-(f), we show the constructed two gain curves of the JPC corresponding to the positive (red) and negative (blue) current pulses measured for different current amplitudes (which induce different flux variation $\Delta\Phi$ in the JRM). As expected, the two solid red and blue gain curves vary with the applied current amplitude. In the case of zero-amplitude current pulses, the gain curves overlap (Fig. \ref{FastFreqSwitching} (d)). As the current amplitude is increased, the center frequencies of the gain curves deviate from each other (Fig. \ref{FastFreqSwitching} (e)) and, consequently, the curves become distinguishable. By further increasing the applied current pulse amplitude, the two gain curves of dynamical bandwidth of $3$ $\operatorname{MHz}$ become separated by $4$ $\operatorname{MHz}$   while attaining power gain of $23$ dB (Fig. \ref{FastFreqSwitching} (f)).

It is worth noting that for a fixed pump frequency $f_P$, such as in our case, the maximum frequency separation between the gain curves that can be achieved using this technique without significantly degrading the gain (e.g., below $20$ dB), is in essence bounded by the smaller bandwidth of the two resonators \cite{JPCreview,Jamp}. This is because both resonance frequencies of the S and I shift in the same direction in response to the applied current pulse, while the amplification condition requires that the sum of the amplified S and I frequencies (residing within the respective bandwidths) stay equal to $f_P$. Consequently, when the shifts in the resonance frequencies become comparable to the bandwidths of the resonators, the amplification condition, set for the fixed-flux case, becomes difficult to satisfy, which in turn leads to a decrease in the gain. 

Another important figure related to time-multiplexed readout of quantum systems is the lower bound on $t_{\mathrm{sw}}$, which is mainly limited by the measurement time needed to obtain a desired readout fidelity. This figure sets an upper bound on the rate at which a time-multiplexed readout can be performed. In the case of qubit-cavity systems, it is quite feasible to jointly optimize their parameters and the JPC's to yield high readout fidelity $>97\%$ with fast pulses $t_{\mathrm{sw}}<400$ ns \cite{QubitJPC,FeedbackJPC}. 

In conclusion, we have realized and measured a JPC device which does not require the use of off-chip, bulky, broadband hybrids and external magnetic coils for its operation. In the new JPC circuit, the pump drive is fed through an on-chip, three-port power divider capacitively coupled to the JRM, and the flux bias is applied through an on-chip flux line. By eliminating the need for microwave hybrids and magnetic coils, we significantly reduce the device footprint without degrading its performance or adding complexity to the fabrication process. In addition, we have utilized the on-chip flux line to demonstrate time-multiplexed amplification of microwave tones that are separated by more than the dynamical bandwidth of the JPC by varying the magnetic flux threading the JRM using fast current pulses. Such a measurement technique can be potentially employed in the readout of pairs of qubit-resonator systems coupled to the same JPC. Furthermore, the new simplified JPC circuit, in which the pump, signal, and idler have separate physical ports, opens the door for the realization of broadband JPCs using impedance engineering techniques \cite{StrongEnvCoupling,JPAimpedanceEng}, and also useful quantum devices such as on-chip circulators and Josephson directional amplifiers \cite{DircJPC,JDA}.

\begin{acknowledgments}
We thank Firat Solgun for fruitful discussions, James Rozen and John Rohrs for helpful technical support, Brian Vlastakis, Srikanth Srinivasan, and William Shanks for valuable programing and software assistance, and Christian Baks for the design and supply of PCBs used in mounting the JPC chips.  
\end{acknowledgments}

\end{document}


\title{Supplemental Material for ``Time-multiplexed amplification in a hybrid-less and coil-less Josephson parametric converter''}
\author{Baleegh Abdo}
\author{Jose M. Chavez-Garcia}
\author{Markus Brink}
\author{George Keefe}
\author{Jerry M. Chow}
\affiliation{IBM T. J. Watson Research Center, Yorktown Heights, New York 10598, USA.}
\date{\today}

\maketitle

\section{The dependence of the resonance frequency, bandwidth, and gain of the JPC on the external magnetic flux}

The signal and idler microstrip resonators of the JPC denoted here as resonators \textit{a} and \textit{b} couple to an inductively shunted JRM as shown in Fig. 1 (c). In addition to serving as a dispersive, nonlinear medium which facilitates three-wave mixing, the inductively shunted JRM functions as a flux-dependent inductor which can be used to tune the resonance frequencies of the device.

 \begin{figure}
 	[h]
 	\begin{center}
 		\includegraphics[
 		width=0.6\textwidth
 		]%
 		{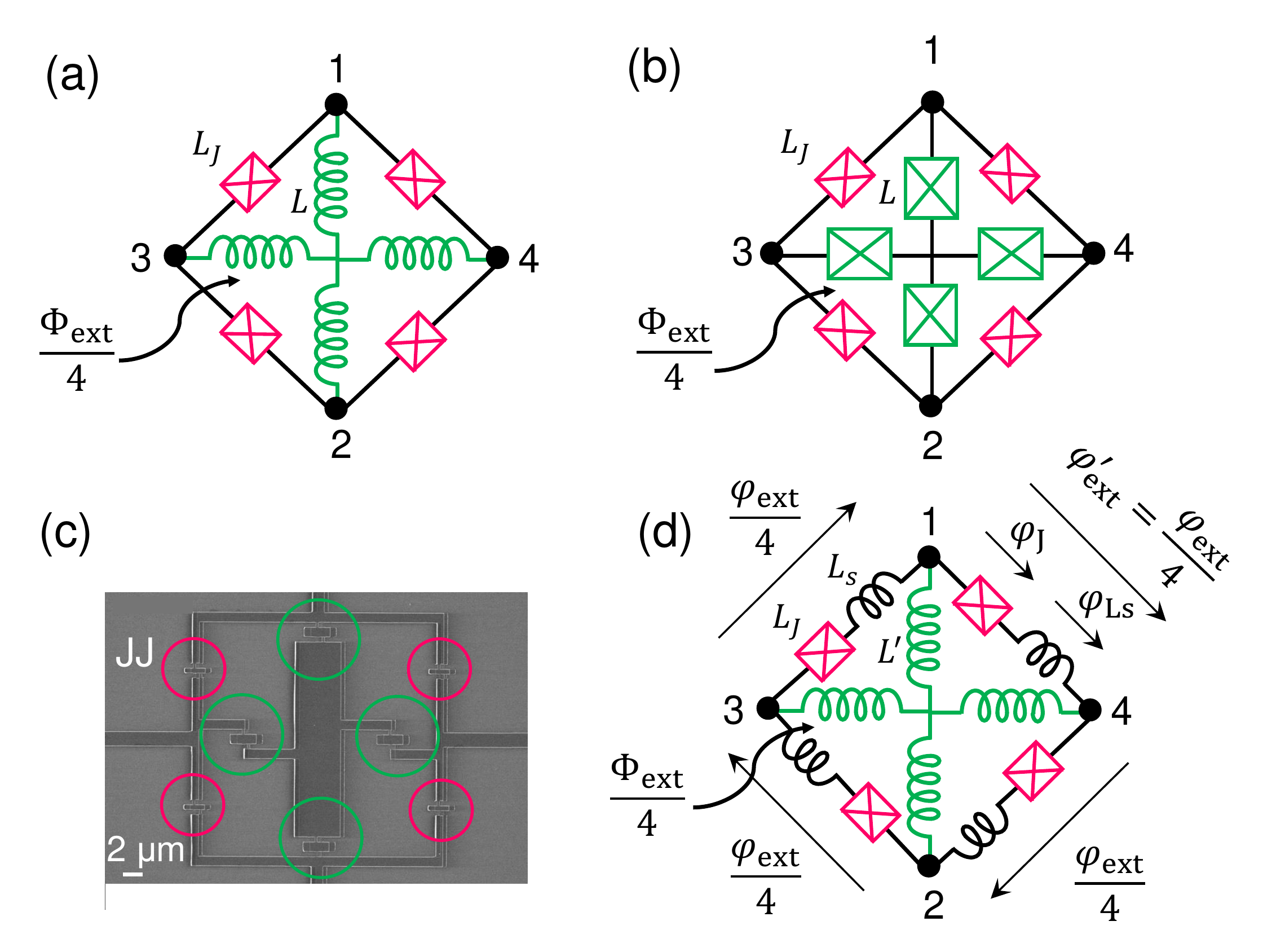}%
 		\caption{(color online). (a) An ideal circuit model of the shunted JRM. It consists of four Josephson junctions arranged in a Wheatstone bridge configuration. Each junction is shunted by two linear inductances denoted $L$. The four inner loops of the JRM are symmetric, and each  covers a quarter of the JRM area. In order to vary the inductance of the JRM, an external magnetic flux is applied to the JRM loop. (b) Shunted JRM in which the linear shunt inductances $L$ inside the loop shown in panel (a) are substituted by an equivalent inductance of large Josephson junctions. (c) A scanning electron micrograph showing an implementation of the shunted JRM which employs large Josephson junctions as shown in the JRM scheme of panel (b). (d) A more realistic circuit model of the shunted JRM of panels (a) and (b), which includes the effect of parasitic series inductance $L_S$ due to the superconducting wires in each arm. The shunt inductances $L^{'}$ represents the sum of the inductance of the large Josephson junction and the superconducting wires inside the JRM loop.}%
 		\label{JRM}%
 	\end{center}
 \end{figure}
 
\begin{figure}
	[h]
	\begin{center}
		\includegraphics[
		width=0.5\textwidth
		]%
		{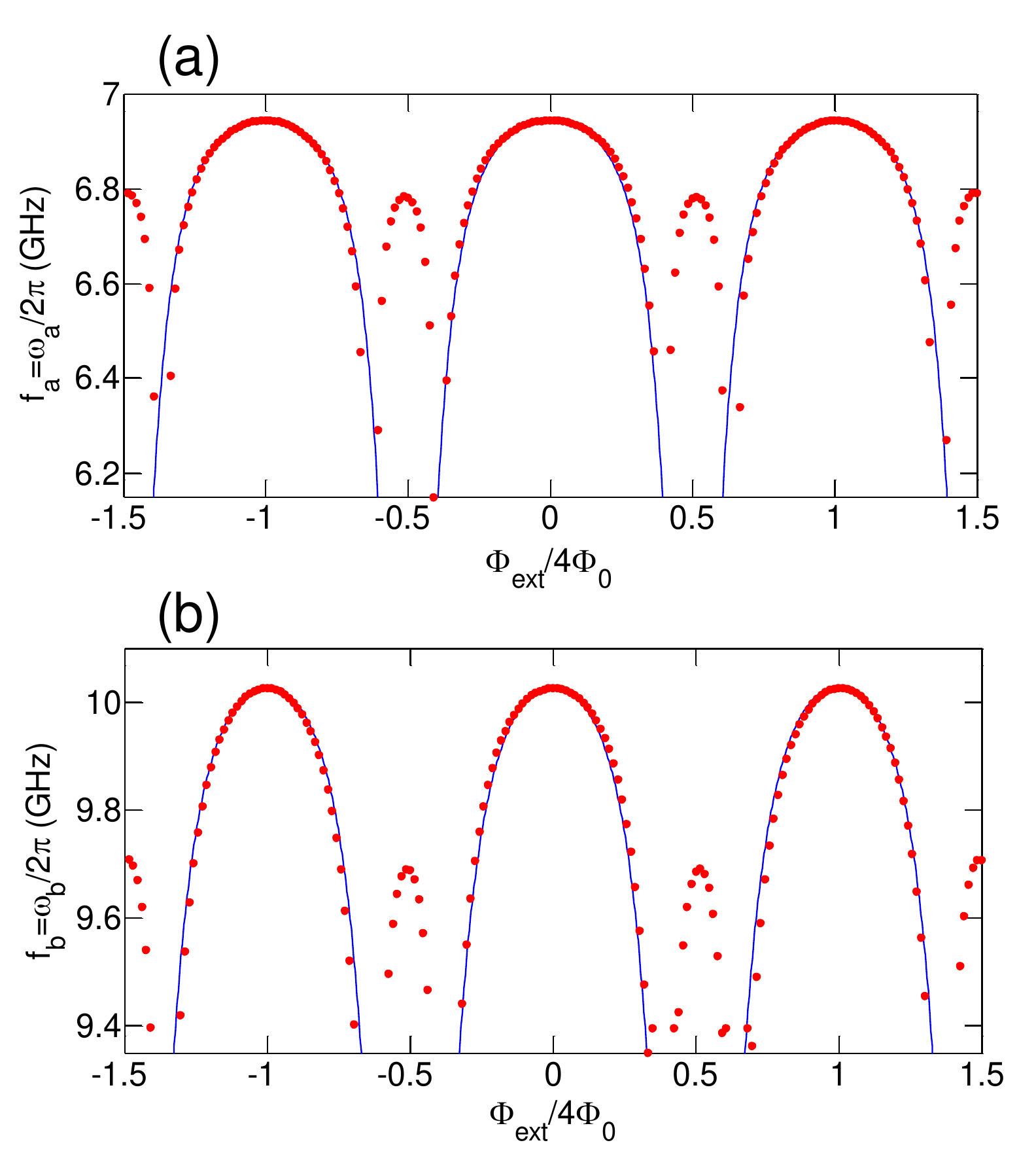}%
		\caption{(color online). Resonance frequency measurement of resonators \textit{a} (panel(a)) and \textit{b} (panel (b)) plotted using filled red circles versus applied external flux threading the inner loops of the JRM. The external magnetic flux is varied using an external coil attached to the JPC package. The solid blue curves are fits to the primary flux lobes of the data calculated using Eq. \ref{w_ab}.}
		\label{ResFluxDep}%
	\end{center}
\end{figure}

 An ideal circuit of the inductively shunted JRM consists of four Josephson junctions arranged in a Wheatstone bridge configuration as shown in Fig. \ref{JRM} (a), where each junction is shunted by two linear inductances denoted $L$. As shown in Ref. \cite{Roch}, for a certain ratio of $L/L_J$ (approximately $1/4<L/L_J<1/2$) where $L_J$ is the inductance of the Josephson junction, such a configuration lifts the hysteretic behavior of the JRM with respect to the applied magnetic flux $\Phi_{\mathrm{ext}}$ threading the loop, therefore enabling the device to operate at a wide range of fluxes in contrast to JPCs with unshunted JRMs which are constrained in their operation to flux bias points in the vicinity of odd-integer multiples of half flux quantum $\Phi_0/2$ \cite{JPCreview,Jamp,JPCnaturePhys} (where $\Phi_0=h/2e$). One straightforward method for realizing the shunt linear inductances is using large Josephosn junctions \cite{FlaviusPhD} as illustrated in Fig. \ref{JRM} (b). The main advantage of using large Josephson junctions versus superconducting wires used in Ref. \cite{Roch} for instance, is that they are easier to scale with respect to the inductance of the outer junctions, and they occupy smaller area. In Fig. \ref{JRM} (c) we show a scanning electron micrograph of a shunted JRM. The Josephson junctions are $\mathrm{Al/AlO_{x}/Al}$ implemented using a bridge-free double angle evaporation technique \cite{BridgeFree}. Figure \ref{JRM} (d) shows a more realistic circuit model of the shunted JRM which accounts for the parasitic series inductance of the superconducting wires connecting the Josephson junctions on the outer loop $L_S$ and on the inner loops where  $L^{'}$ represents the shunt inductance due to the large Josepshon junction and the superconducting wires. Using the notations of Fig. \ref{JRM} (d), the Josephson inductance $L_{J}$ can be written as 

\begin{equation}
L_{J}(\varphi_{\mathrm{J}})=\frac{L_{J0}}{cos(\varphi_{\mathrm{J}})},
\label{L_J}
\end{equation}         

where $\varphi_J$ is the phase difference across the Josephson junction, $L_{J0}=\phi_0/I_0$, where $\phi_{0}=\Phi_0/2\pi$ and $I_0$ is the critical current. Due to the JRM symmetry, the phase difference across each branch of the outer loop of the JRM given by $\varphi_{\mathrm{ext}}^{'}=\varphi_{\mathrm{J}}+\varphi_{L_S}$ can be expressed as $\varphi_{\mathrm{ext}}^{'}=\varphi_{\mathrm{ext}}/4$, where $\varphi_{L_S}$ is the phase difference across the series inductance $L_S$ and $\varphi_{\mathrm{ext}}\equiv\Phi_{\mathrm{ext}}/\phi_{0}$. In this model, we are limiting ourselves to flux configurations in which the net circulating current in the JRM flows on the outer loop with no dc current flowing in the shunt inductance \cite{Roch,FlaviusPhD}. Since the same current flows in the JJ and the series inductance, the following relation holds 

\begin{equation}
E_{L_S}\varphi_{L_S}=E_J\sin(\varphi_{\mathrm{J}}),
\end{equation}

where $E_{L_S}=\phi_0^{2}/L_S$ and $E_J=\phi_0^{2}/L_{J0}$.

In our case $E_J\ll E_{L_S}$ ($L_S\ll L_{J0}$), thus $\varphi_{\mathrm{J}}$ can be expanded as \cite{Roch}

\begin{align}
\varphi_{\mathrm{J}} & =\varphi_{\mathrm{ext}}^{'}-\frac{E_J}{E_{L_S}}sin(\varphi_{\mathrm{J}}),\\
& 
\approx\varphi_{\mathrm{ext}}^{'}-\frac{E_J}{E_{L_S}}sin(\varphi_{\mathrm{ext}}^{'}).
\end{align}

The equivalent inductance of the shunted JRM as seen by `X' and `Y' excitations of the JRM is given by 
\begin{equation}
L_{X,Y}=2(L_J+L_S) || 2(L_J+L_S) || 2L^{'},
\end{equation}

which can be written as    

\begin{equation}
L_{X,Y}(\varphi_{\mathrm{ext}}^{'})=\dfrac{2L^{'}(L_S+L_J(\varphi_{\mathrm{ext}}^{'}))}{2L^{'}+L_S+L_J(\varphi_{\mathrm{ext}}^{'})}.
\label{L_XY}
\end{equation}

Around the fundamental resonance, the resonators \textit{a} and \textit{b} of characteristic impedance $Z_{a,b}$, can be modeled as lumped-element resonators with flux-dependent angular resonance frequency   

\begin{equation}
\omega_{a,b}\left(\varphi_{\mathrm{ext}}^{'}\right)=\dfrac{1}{\sqrt{C_{a,b}L_{_{a,b}}\left(\varphi_{\mathrm{ext}}^{'} \right) }},
\label{w_ab}
\end{equation}

where $L_{a,b}(\varphi_{\mathrm{ext}}^{'})\equiv L_{a,b}^{0}+L_{X,Y}(\varphi_{\mathrm{ext}}^{'})$, $C_{a,b}=\pi/(2Z_{a,b}\omega_{a,b}^{\mathrm{max}})$, $L_{a,b}(0)=2Z_{a,b}/(\pi\omega_{a,b}^{\mathrm{max}})$, $L_{a,b}^{0}\equiv L_{a,b}(0)-L_{X,Y}(0)$,  and $\omega_{a,b}^{\mathrm{max}}$ (e.g., $\omega_{a,b}(0)$) is the maximum angular frequency of the resonators. In Fig. \ref{ResFluxDep} we exhibit a resonance frequency measurement of resonators \textit{a} and \textit{b} plotted using filled red circles versus applied external flux threading the inner loops of the JRM (quarter of the JRM loop). The solid blue curves are fits to the data (i.e., the primary flux lobes which support amplification), calculated using Eq. \ref{w_ab}, where $f_{a,b}=\omega_{a,b}/2\pi$.

By using the measured maximum resonance frequencies $f_a^{\mathrm{max}}=\omega_{a}^{\mathrm{max}}/2\pi=6.945$ $\operatorname{GHz}$, $f_b^{\mathrm{max}}=\omega_{b}^{\mathrm{max}}/2\pi=10.03$ $\operatorname{GHz}$ and our estimate of the critical current $I_0=3$ $\mu\operatorname{A}$ of the Josephson junctions, we extract the values of the following device parameters, $Z_{a,b}=55$ $\operatorname{Ohm}$, $C_a=0.65$ p$\operatorname{F}$, $C_b=0.45$ p$\operatorname{F}$, $L_{a}^{0}=0.75$ n$\operatorname{H}$, $L_{b}^{0}=0.51$ n$\operatorname{H}$, $L_{J0}=110$ p$\operatorname{H}$, $L_{S}=10.6$ p$\operatorname{H}$, and $L^{'}=44.2$ p$\operatorname{H}$. 

To calculate the dependence of the resonator bandwidth on the external flux we start with the external quality factor for the fundamental modes of the resonators \textit{a} and \textit{b}, given by

\begin{equation}
Q_{\mathrm{ext}}^{a,b}=\left( \frac{C_{a,b}}{C_c^{a,b}}\right) ^{2}\dfrac{1}{\omega_{a,b}C_{a,b}Z_{0}}, 
\end{equation}

where $C_c^{a,b}$ are the coupling capacitors which couple between the resonators \textit{a} and \textit{b} and the external feedlines of characteristic impedance $Z_0=50$ $\operatorname{Ohm}$ which carry the ingoing and outgoing signals. 
Since the JPC resonators are strongly coupled (the internal quality factor $Q_{\mathrm{int}}^{a,b}$ is much larger than $Q_{\mathrm{ext}}^{a,b}$), the total quality factor $Q_{\mathrm{tot}}^{a,b}$ given by $1/Q_{\mathrm{tot}}^{a,b}=1/Q_{\mathrm{int}}^{a,b}+1/Q_{\mathrm{int}}^{a,b}$ is limited by $Q_{\mathrm{ext}}^{a,b}$, i.e., $Q_{\mathrm{tot}}^{a,b}\cong Q_{\mathrm{ext}}^{a,b}$. Hence, we obtain a flux-dependent resonator bandwidth which reads 

\begin{align}
\gamma_{a,b}(\varphi_{\mathrm{ext}}^{'})) &
=\dfrac{\omega_{a,b}}{Q_{\mathrm{ext}}^{a,b}}\\
& 
=\left( \frac{C_c^{a,b}}{C_{a,b}}\right) ^{2}\dfrac{Z_{0}}{L_{a,b}(\varphi_{\mathrm{ext}}^{'})}.
\label{gammaAB}
\end{align}

By substituting the dependence of the device resonance frequency and bandwidth (Eqs. (\ref{w_ab}), and (\ref{gammaAB})) on the applied magnetic flux into the theoretical expression for the reflection parameter of the JPC under the stiff pump approximation we get \cite{JPCreview}    

\begin{equation}
r_{aa}[\omega_{1}] =\dfrac{\chi_{a}^{-1*}[\omega_{1}]\chi_{b}^{-1*}[\omega_{2}]+\left|\rho \right| ^{2}}{\chi_{a}^{-1}[\omega_{1}]\chi_{b}^{-1*}[\omega_{2}]-\left|\rho \right| ^{2}}, 
\label{raa}
\end{equation}

where

\begin{equation}
\chi_{a}^{-1}[\omega_{1}]=1-2i\dfrac{\omega_{1}-\omega_{a}(\varphi_{\mathrm{ext}}^{'})}{\gamma_{a}(\varphi_{\mathrm{ext}}^{'})}, 
\end{equation}

\begin{equation}
\chi_{b}^{-1}[\omega_{2}]=1-2i\dfrac{\omega_{2}-\omega_{b}(\varphi_{\mathrm{ext}}^{'})}{\gamma_{b}(\varphi_{\mathrm{ext}}^{'})}, 
\label{Chibw2}
\end{equation}

are the inverses of the bare response functions of modes \textit{a} and \textit{b} which depend linearly on the signal frequency at $\omega_{1}$ or $\omega_{2}$ (which lie within the device bandwidth), and $\rho$ is a dimensionless pump amplitude whose absolute value varies in the range $0\leq\left| \rho \right| <1$ (where $\left| \rho \right|=0$ correspond to the case of no applied pump, i.e., no amplification, and $\left| \rho \right|\longrightarrow1^{-}$ to the case of large amplification).

Furthermore, since the pump frequency is commonly set to the frequency sum of the resonance frequencies of resonators \textit{a} and \textit{b} and the tone frequencies $\omega_{1}$ and $\omega_{2}$ are interrelated through the applied pump frequency, the following relation holds $\omega_{p}=\omega_{a}+\omega_{b}=\omega_{1}+\omega_{2}$. By applying this relation we can rewrite Eq. \ref{Chibw2} as

\begin{equation}
\chi_{b}^{-1}[\omega_{1}]=1-2i\dfrac{\omega_{a}(\varphi_{\mathrm{ext}}^{'})-\omega_{1}}{\gamma_{b}(\varphi_{\mathrm{ext}}^{'})}. 
\end{equation}

Using Eq. \ref{raa} we obtain the device gain versus frequency and applied magnetic flux given by 

\begin{equation}
G[\omega_{1}]=\left|r_{aa}[\omega_{1}] \right|^{2}. 
\label{G}
\end{equation}

In the case discussed in the paper where the switching rate in the applied magnetic flux is much smaller than the dynamical bandwidth of the device, i.e., $1/t_{\mathrm{sw}}\ll B$, we can express the applied magnetic flux following each switching event as a small fixed deviation from the flux bias point $\Phi_{\mathrm{b}}$

\begin{equation}
\Phi_{\mathrm{ext}}^{\pm}(\pm\delta)=\Phi_{\mathrm{b}}(1\pm\delta), 
\label{PhiExtPm}
\end{equation}

where $|\delta|\ll 1$. Thus, the change in the flux threading the JRM loop can be expressed as $\Delta\Phi=\pm\Phi_{\mathrm{b}}\delta$. 

To reproduce the theoretical gain curves in Fig. 4, Eq. \ref{G} is used with the parameters $\Phi_{\mathrm{b}}=0.64\Phi_{\mathrm{0}}$ and $\left|\rho \right| ^{2}=0.92$.

Please note that in this section we only consider the case of port (\textit{a}), but a similar result can be derived for port (\textit{b}) as well. 

\section{Injecting the pump drive through the power divider}

\begin{figure}
	[tb]
	\begin{center}
		\includegraphics[
		width=0.7\columnwidth 
		]%
		{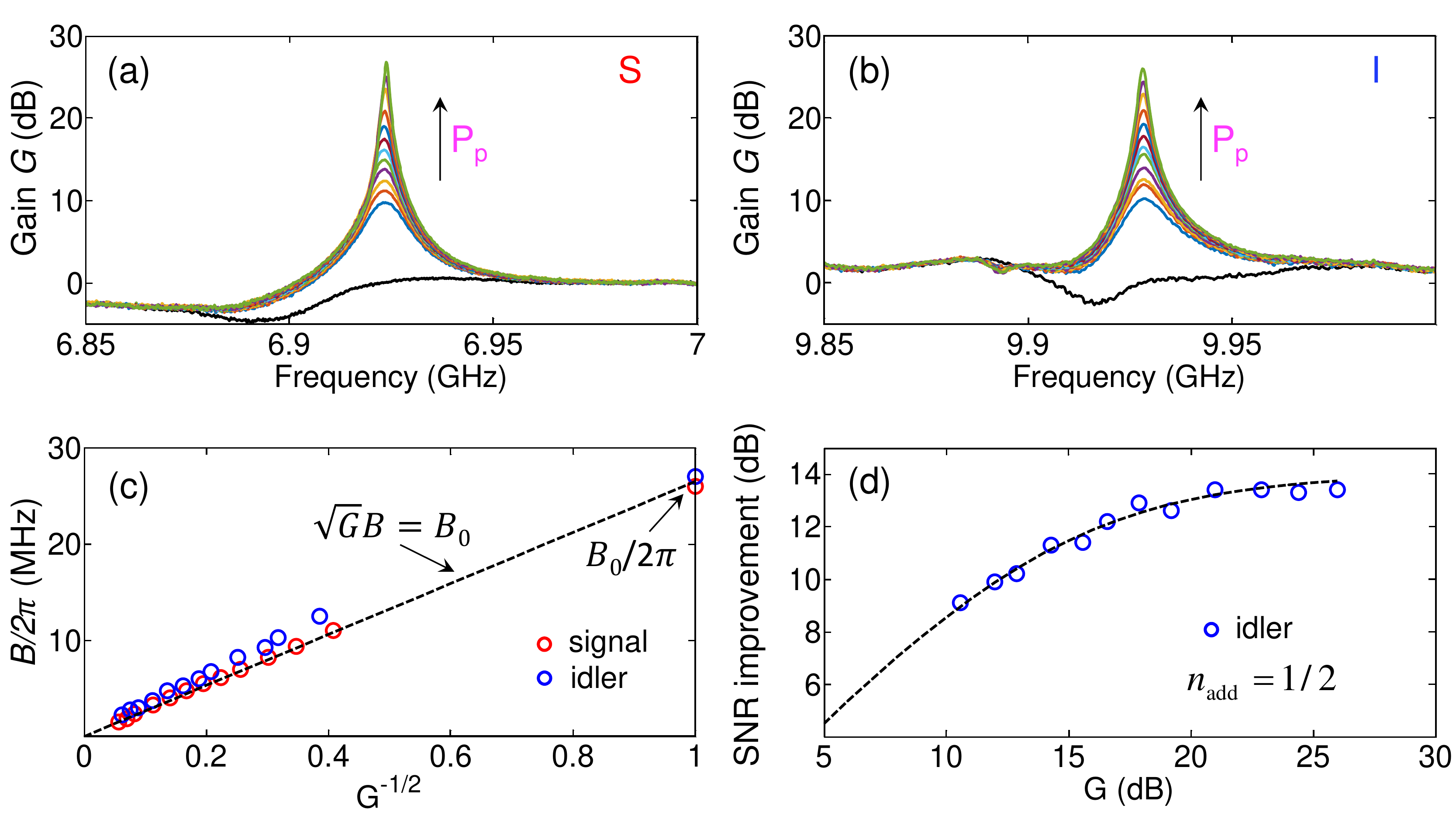}
		\caption{(color online). Gain curves versus signal (a) and idler (b) frequency measured at a fixed flux bias point. The different gain curves in subplots (a) and (b) correspond to different pump powers. (c) Extracted dynamical bandwidth $B/2\pi$ of the signal (red circles) and idler (blue circles) versus $G^{-1/2}$. The dashed black line represents the amplitude-gain bandwidth product relation. (d) Signal to noise ratio (SNR) improvement measurement (blue circles) versus gain taken on the idler port on resonance. The black dashed line corresponds to the calculated SNR improvement of the JPC versus gain. The fit parameter $n_{\mathrm{add}}=1/2$ indicates that the new device operates near the quantum limit.  
		}
		\label{Gain}
	\end{center}
\end{figure}

In Fig. \ref{Gain}, we show typical measurement results of the new JPC device operated in the amplification mode. These results show that feeding the pump through the power divider achieves comparable performance  to state-of-the-art JPCs which employ hybrids. In these measurements, the JPC is flux biased to a center frequency of $f_S=6.924$ $\operatorname{GHz}$ and $f_I=9.928$ $\operatorname{GHz}$ using an external magnetic coil attached to the JPC package. The pump tone is applied through the power divider port at frequency $f_p=16.852$ $\operatorname{GHz}$. Figure \ref{Gain} (a) and (b) exhibit several gain curves measured in reflection on the signal and idler ports respectively for varying pump powers. Both panels show that the device power gain $G$ increases with the applied pump power and gains in excess of $20$ dB required for high-fidelity readout can be achieved on both ports. The bottom black curves in both panels represent the JPC amplitude response without pump. The small dip at around $6.9$ $\operatorname{GHz}$ and $9.91$ $\operatorname{GHz}$ of about $1$-$2$ dB implies that a certain amount of power leaks out from the S and I resonators through the power divider and flux line. However, this figure can be improved using a more elaborate microwave design or filtering. In Fig. \ref{Gain} (c), we plot the dynamical  bandwidths of the JPC $B/2\pi$ (i.e., $-3$ dB points from the maximum gain) extracted from the gain curves of the S (Fig. \ref{Gain} (a)) and I (Fig. \ref{Gain} (b)) ports versus $G^{-1/2}$. The red (blue) circles correspond to the measured $B$ on the S (I) port. The dashed black line corresponds to the amplitude-gain bandwidth product relation $B\sqrt{G}=B_{0}$ characteristic of Josephson parametric amplifiers employing resonators in the limit of large gains \cite{JPCreview,NoiseAmplReview}, where $B_{0}=2\gamma_{a}\gamma_{b}/(\gamma_{a}+\gamma_{b})$. In Fig. \ref{Gain} (d) we measure the signal to noise (SNR) improvement of the output chain on the idler side given by the ratio $G/G_{N}$ versus gain, where $G_{N}$ is the rise in the noise floor of the output line due to the JPC. By fitting this measurement to the theoretical expression $G/G_{N}=T_{N}/(T_{N}G^{-1}+T_{Q}(1/2+n_{\mathrm{add}}))$, where $T_{Q}=hf_{I}/k_{B}$, and substituting the noise temperature of the output line $T_{N}=12\pm1$ $\operatorname{K}$, we get an added input noise photon value of $n_{\mathrm{add}}=1/2$ in good agreement with the data, implying that the device operates near the quantum limit in accordance with other JPCs \cite{JPCnature}.

\begin{figure}
	[h!]
	\begin{center}
		\includegraphics[
		width=0.4\textwidth
		]%
		{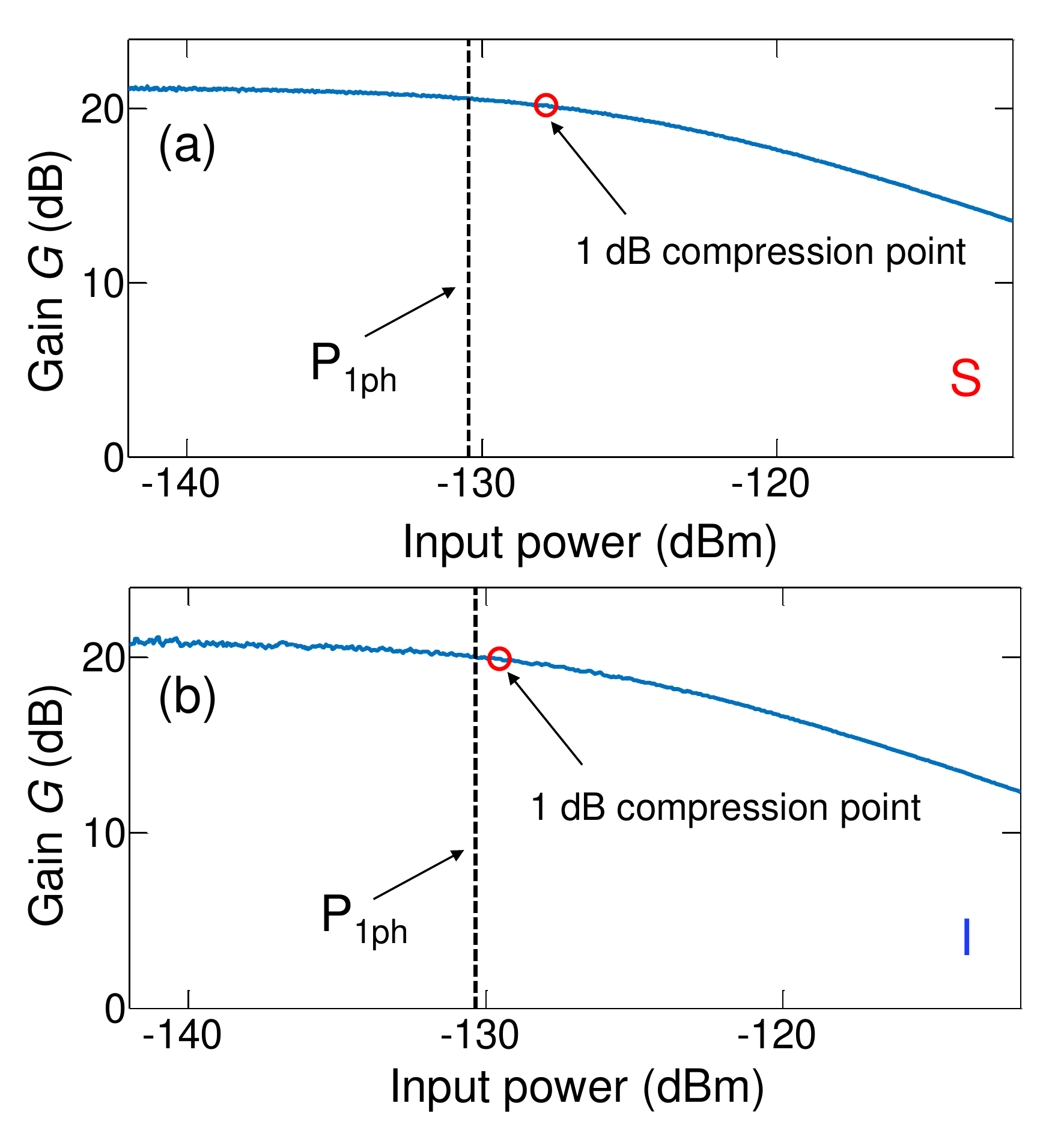}%
		\caption{(color online). Plots (a) and (b) correspond to maximum input power measurements at $21$ dB of gain on resonance taken on the S and I ports at $f_S=6.924$ $\operatorname{GHz}$ and $f_I=9.928$ $\operatorname{GHz}$ respectively. In both measurements, the pump drive is applied through the on-chip power divider at frequency $f_P=16.852$ $\operatorname{GHz}$. The black dashed vertical line indicates the input power of one photon at the signal or idler frequency per inverse dynamical bandwidth of the device. The red open circle marks the $1$ dB compression point in each measurement.}
		\label{DRPD}%
	\end{center}
\end{figure}

In Fig. \ref{DRPD}, we show in panels (a) and (b) a maximum input power measurement taken on the S and I ports respectively on resonance (for the same measurement parameters of Fig. \ref{Gain}). In both measurements, the device gain is set to about $G=21$ dB and the input power of the applied S and I tones is varied. The red open circle in each graph marks the $1$ dB compression point which corresponds to the input power at which the device gain drops by $1$ dB and the amplifier starts to saturate. The black dashed vertical line indicates the input power of one photon at the signal or idler frequency per inverse dynamical bandwidth of the device.    

\section{The dependence of the resonance frequency of the JPC on the dc current applied to the flux line}

\begin{figure}
	[h!]
	\begin{center}
		\includegraphics[
		width=0.3\textwidth
		]%
		{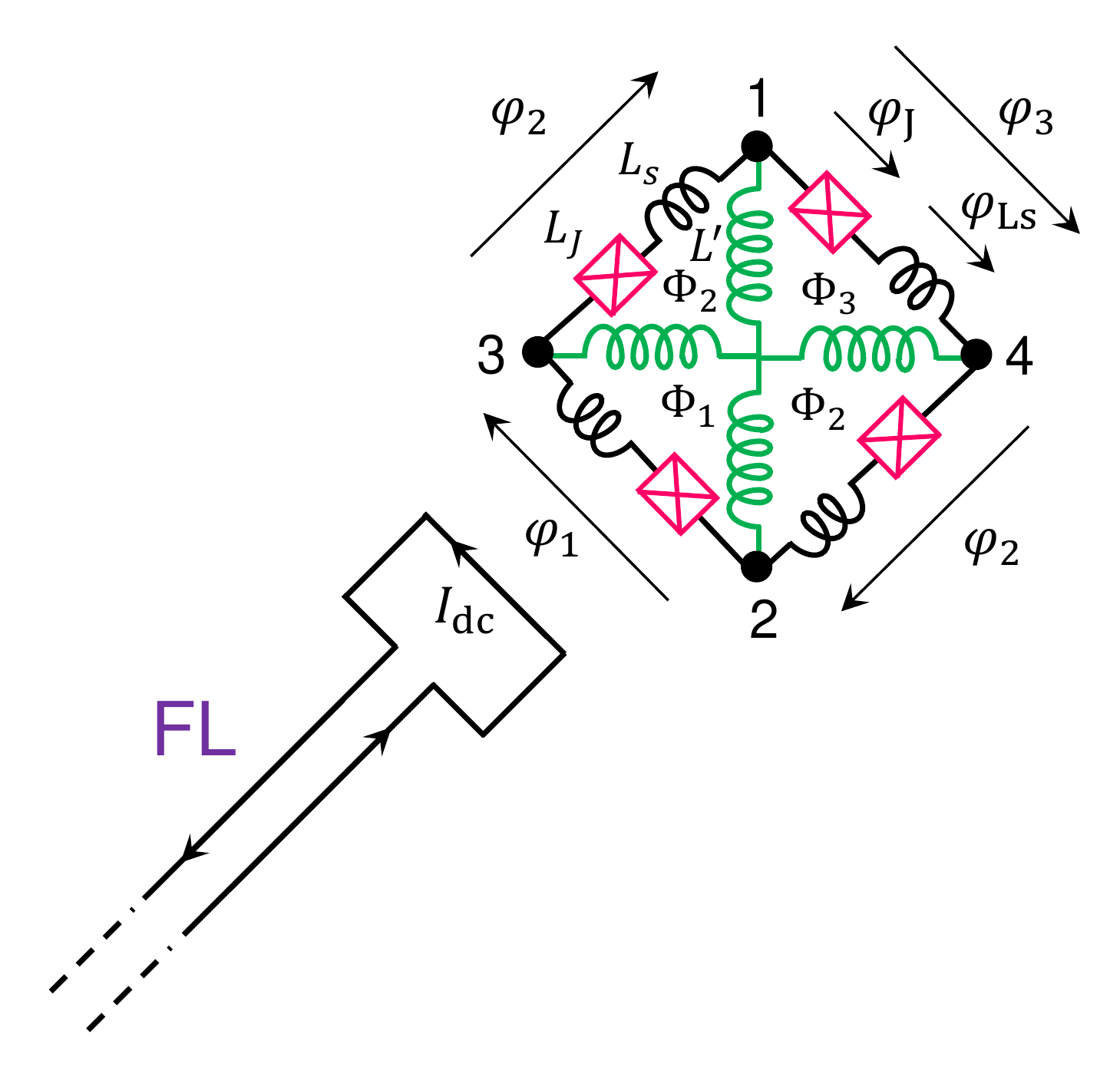}%
		\caption{(color online). A drawing (not to scale) featuring the shunted JRM and the expected effect of a dc current applied to the on-chip flux line. In this picture, the induced magnetic fluxes in the JRM quarters (the four equal inner loops) is expected to vary depending on the relative distance between the quarters and the flux line, which would in turn vary the phase difference across each outer arm of the JRM.}
		\label{DCFLJRM}%
	\end{center}
\end{figure}

\begin{figure}
	[h!]
	\begin{center}
		\includegraphics[
		width=0.4\textwidth
		]%
		{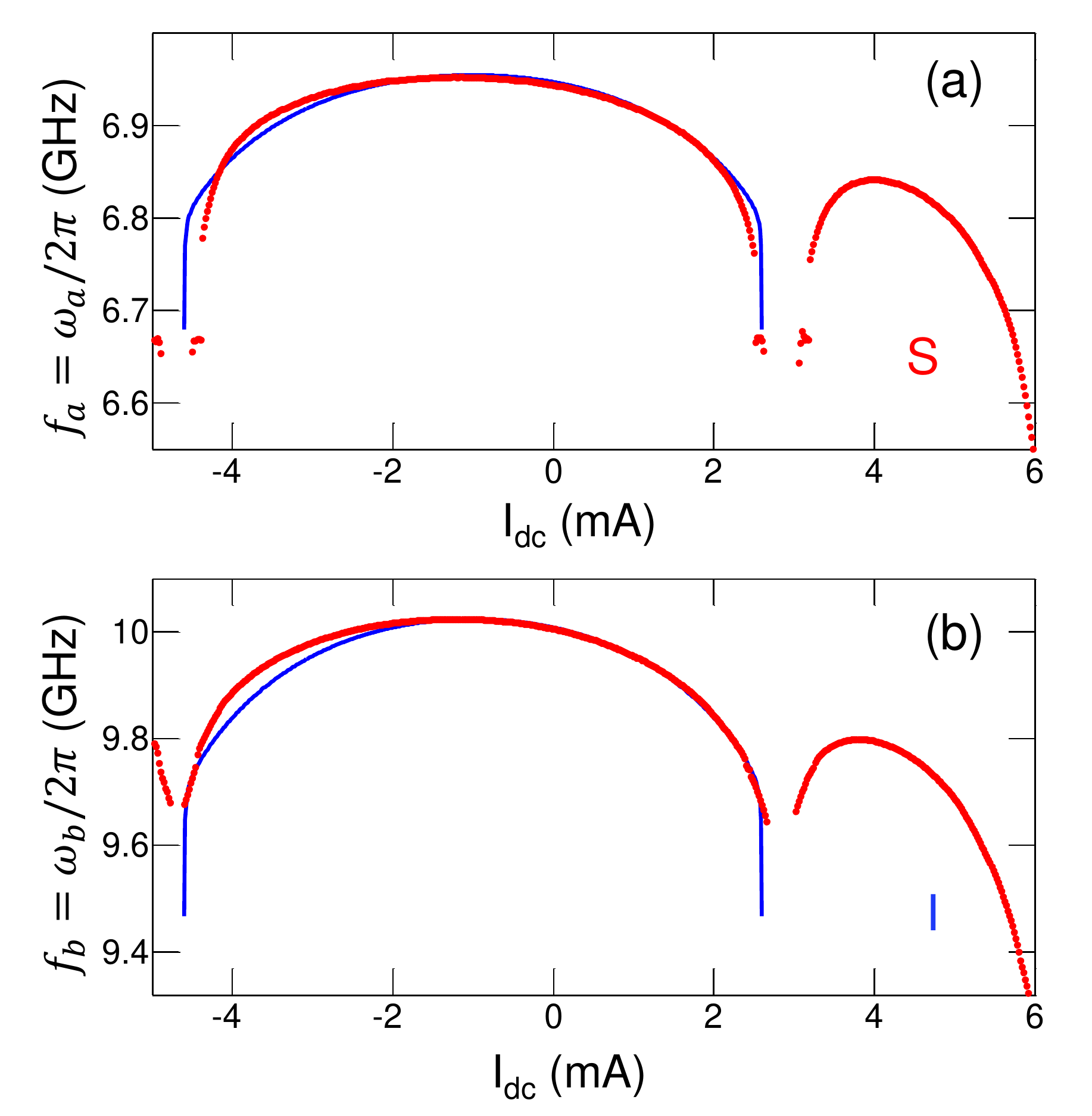}%
		\caption{(color online). Resonance frequency measurement of resonators \textit{a} (panel(a)) and \textit{b} (panel (b)) plotted using filled red circles versus dc current applied through the flux line. The solid blue curves represent fits to the primary flux lobe based on the modified model presented in section III with the parameters, $M=0.57$ p$\operatorname{H}$, $\alpha_{1}=0.95$, and $\alpha_{2}=0.87$.}
		\label{DCFLMeas}%
	\end{center}
\end{figure}

One important difference between the external magnetic coil shown in Fig. 1 (d) and the on-chip flux line shown in Fig. 1 (f) is that the axis of the magnetic coil coincides with the center of the JRM, whereas the on-chip flux line is positioned at the side of the JRM as illustrated in the drawing of Fig. \ref{DCFLJRM}. As a result of this slight asymmetry it is expected that the induced magnetic flux in each quarter of the JRM to be slightly different. In particular, we expect that the two JRM quarters defined by nodes 1 and 3 and 2 and 4 (see Fig. \ref{DCFLJRM} ) to receive approximately equal flux $\Phi_2$ due to their equal distance from the edge of the flux line, which in turn is slightly smaller than the flux  $\Phi_1$ threading the closest JRM quarter defined by nodes 2 and 3, and slightly larger than the flux $\Phi_3$ threading the farthest quarter defined by nodes 1 and 4. 

Defining the phase difference across each branch of the outer loop in terms of the flux threading each quarter  $\varphi_{\mathrm{1,2,3}}\equiv\Phi_{\mathrm{1,2,3}}/\phi_{0}$, and requiring that the sum of the fluxes to be equal to the applied external flux induced by the dc current flowing in the flux line  $\Phi_{\mathrm{ext}}=\Phi_{\mathrm{1}}+2\Phi_{\mathrm{2}}+\Phi_{\mathrm{3}}$, we can write the Josephson inductance in each branch $i=1,2,3$ as      

\begin{equation}
L_{J,i}(\varphi_{\mathrm{J,i}})=\frac{L_{J0}}{cos(\varphi_{\mathrm{J,i}})}.
\label{L_Ji}
\end{equation}    

By further using the approximation $\varphi_{\mathrm{J,i}}\approx\varphi_{\mathrm{i}}-\frac{E_J}{E_{L_S}}sin(\varphi_{\mathrm{i}})$, which links the phase difference across the junction to the total phase across the branch, and neglecting the small dc currents that might flow in the inner loops due to the asymmetry, we can express the total JRM inductance seen by the `X' or `Y' modes as  

\begin{equation}
L_{X,Y}=(L^{'}_{J,2}+L^{'}_{J,3}) || (L^{'}_{J,2}+L^{'}_{J,1}) || 2L^{'},
\label{L_XYFL}
\end{equation}

where $L^{'}_{J,i} \equiv L_{J,i}+L_S$. Furthermore, since the fluxes $\Phi_{\mathrm{i}}$ are proportional to each other and to the applied dc current in the flux line $I_{\mathrm{dc}}$, we can express the external flux $\Phi_{\mathrm{ext}}$ in two forms, 1) $\Phi_{\mathrm{ext}}=(1+2\alpha_1+\alpha_2)\Phi_{\mathrm{1}}$, where the asymmetry parameters $0<\alpha_{1,2}\leq1$ are defined as $\Phi_{\mathrm{2,3}}\equiv\alpha_{1,2}\Phi_{\mathrm{1}}$, and 2) $\Phi_{\mathrm{ext}}=MI_{\mathrm{dc}}$, where $M$ is an effective mutual inductance between the flux line and the JRM. 

By using these relations and substituting Eq. \ref{L_XYFL} in Eq. \ref{w_ab}, while using the device parameters extracted from the theory fits of Fig. \ref{ResFluxDep}, we can calculate the resonance frequency dependence of resonators \textit{a} and \textit{b} on the dc current $I_{\mathrm{dc}}$ applied through the flux line. 

In Fig. \ref{DCFLMeas}, we show a resonance frequency measurement (the red filled circles) taken on the S and I ports for resonators \textit{a} (panel (a)) and \textit{b} (panel (b)) as a function of the dc current applied to the flux line. Both measured data on the S and I ports exhibit one primary flux lobe which supports amplification and one secondary which does not. In this measurement, we limited the scan to one lobe each, since we used a normal metal wire to connect the dc connector at the mixing chamber to the on-chip flux line. This normal connection caused the mixing chamber to heat up in the range $50-100$ m$\operatorname{K}$ beyond $\pm2.5$ m$\operatorname{A}$,  therefore limiting the maximum dc current that can be applied to the device. The solid blue curves represent fits to the primary flux lobe based on the modified model presented in this section. It is important to note that the measured data (and the corresponding fits) is off-centered around zero dc current, most likely due to the presence of a non-zero background magnetic flux inside the magnetic shield can. As can be seen from the plots the theory fits show a relatively good quantitative agreement with the data. As to the slight asymmetry that appears in the measured primary flux lobe around its center (corresponding to the maximum resonance frequency), one possible explanation might be due to enhanced circulating currents in the shunt inductances for large flux biases (which the model does not account for).        
	
\section{Injecting the pump drive through the flux line}

\begin{figure}
	[h!]
	\begin{center}
		\includegraphics[
		width=0.4\textwidth
		]%
		{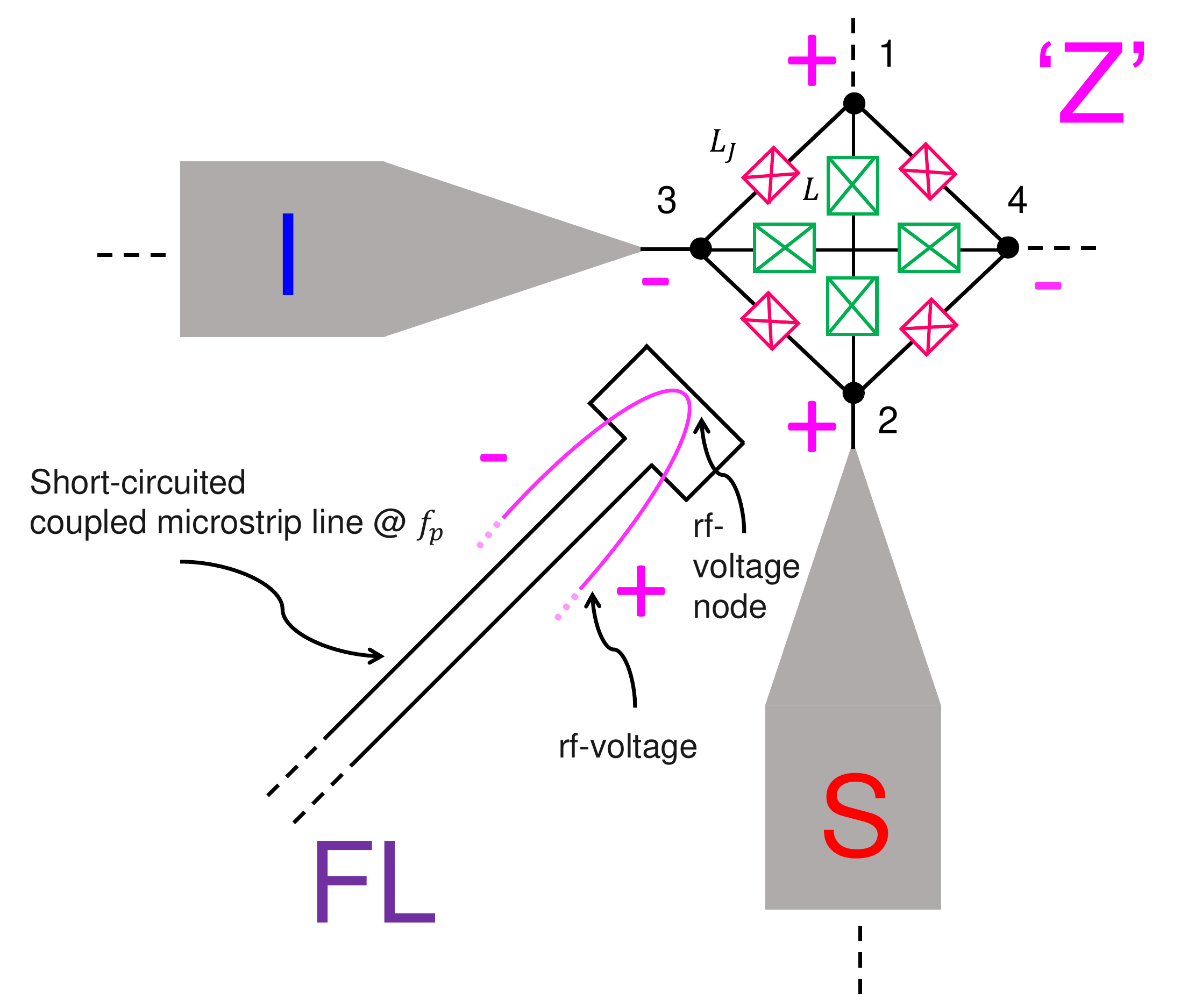}%
		\caption{(color online). A drawing (not to scale) featuring the shunted JRM and the on-chip flux line section in its vicinity. The gray regions connected to nodes $2$ and $3$ of the JRM represent sections of resonators \textit{a} and \textit{b} which support the differential modes S and I of the JRM. At microwave frequencies applicable for the pump drive in the range of $16$-$17$ $\operatorname{GHz}$, the flux line which is implemented in the form of two $5$ $\mu$$\operatorname{m}$ wide strips separated by a $5$ $\mu$$\operatorname{m}$ gap and shorted together near the JRM through a small open loop serves as a short-circuited coupled microstrip line, which supports a differential rf-voltage waveform with a node at the short. Such a differential rf-voltage which capacitively couples to two adjacent nodes of the JRM (e.g., $2$ and $3$) gives rise to a common mode excitation of the JRM `Z' as illustrated in the drawing.}
		\label{PumpFluxLine}%
	\end{center}
\end{figure}

\begin{figure}
	[h!]
	\begin{center}
		\includegraphics[
		width=0.7\textwidth
		]%
		{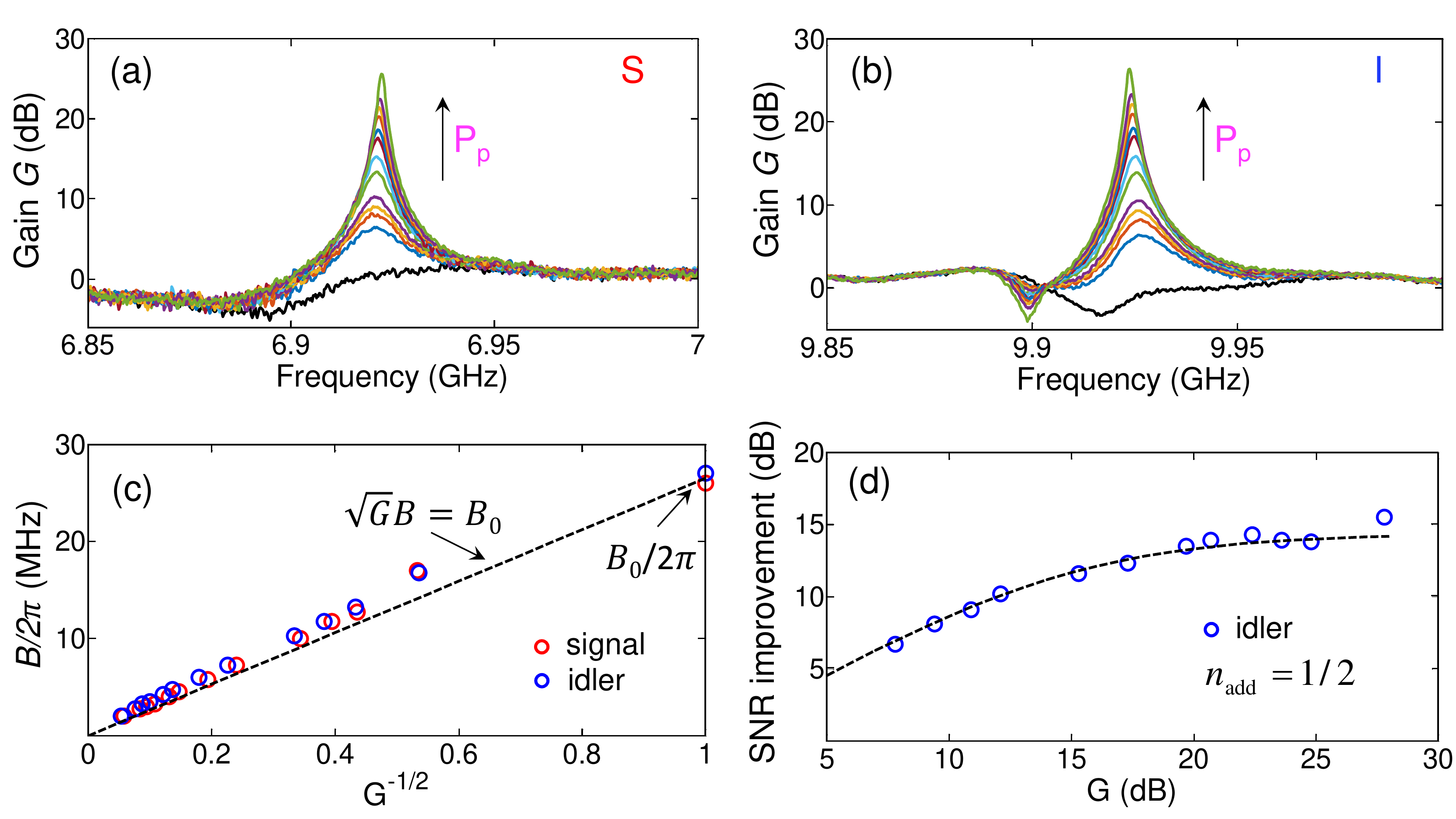}%
		\caption{(color online). Gain curves versus signal (a) and idler (b) frequency measured at a fixed flux bias point. The pump drive is fed to the device through the on-chip flux-line at frequency $f_P=16.846$ $\operatorname{GHz}$. The different gain curves in subplots (a) and (b) correspond to different pump powers. (c) Extracted dynamical bandwidth $B/2\pi$ of the signal (red circles) and idler (blue circles) versus $G^{-1/2}$. The dashed black line represents the amplitude-gain bandwidth product relation which characterizes Josephson parametric amplifiers in the limit of large gain. (d) Signal to noise ratio (SNR) improvement measurement (blue circles) versus gain taken on the idler port on resonance. The black dashed line corresponds to the calculated SNR improvement of the JPC versus gain. The fit parameter $n_{\mathrm{add}}=1/2$ indicates that the new device operates near the quantum limit.}
		\label{GainFL}%
	\end{center}
\end{figure}

\begin{figure}
	[h!]
	\begin{center}
		\includegraphics[
		width=0.4\textwidth
		]%
		{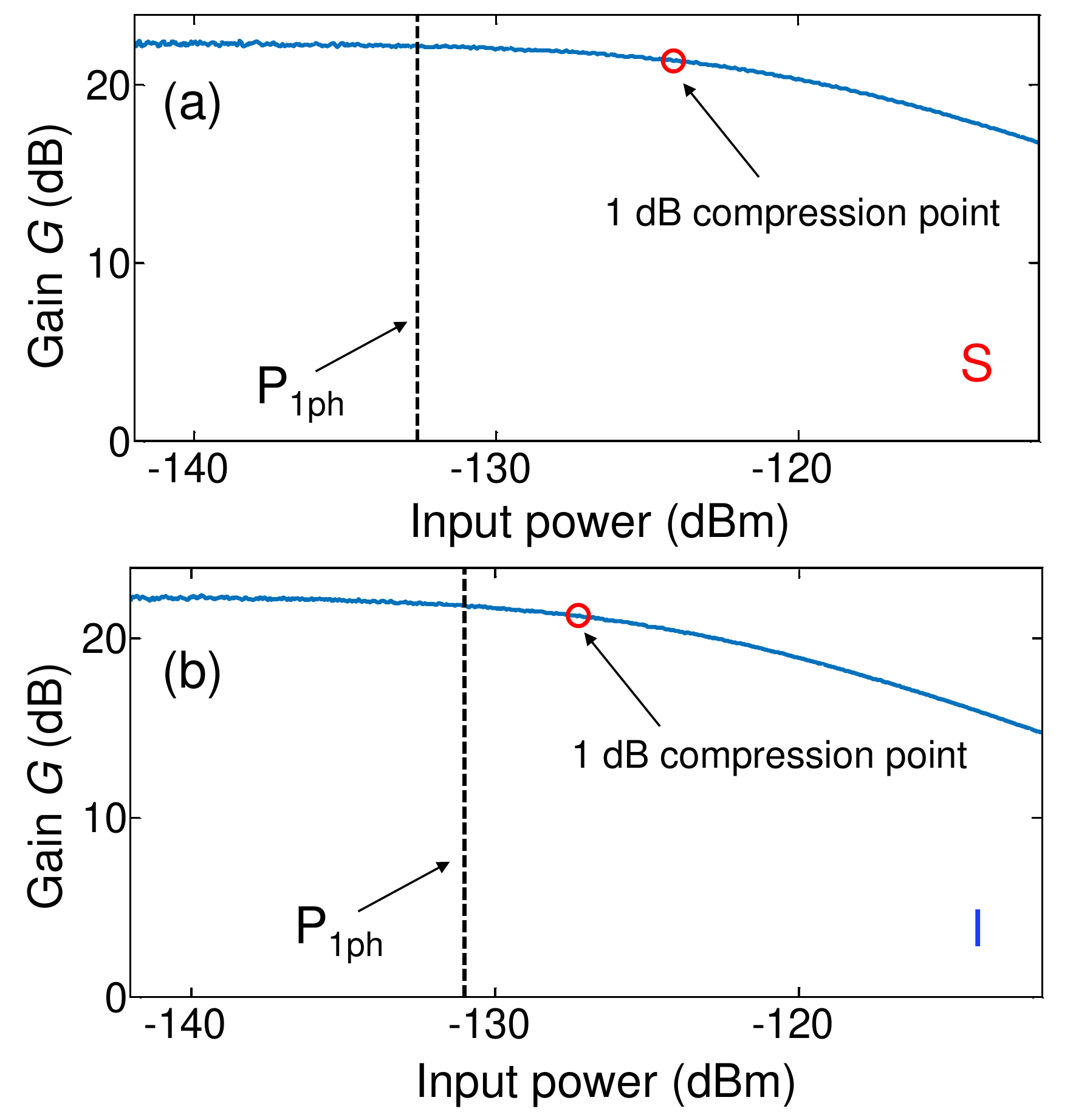}%
		\caption{(color online). Plots (a) and (b) correspond to maximum input power measurements at $23$ dB of gain on resonance taken on the S and I ports at $f_S=6.924$ $\operatorname{GHz}$ and $f_I=9.937$ $\operatorname{GHz}$ respectively. In both measurements, the on-chip flux line is used in order to flux-bias the JRM and apply the pump drive at frequency $f_P=16.861$ $\operatorname{GHz}$ to the device. The dc current inducing the flux bias and the microwave pump drive are input into the flux line using a bias tee. The black dashed vertical line indicates the input power of one photon at the signal or idler frequency per inverse dynamical bandwidth of the device. The red open circle marks the $1$ dB compression point in each measurement.}
		\label{DRFL}%
	\end{center}
\end{figure}

In this section, we show yet another method that can be used in order to inject the pump drive into the JPC without employing any hybrids. Similar to the idea of injecting the pump drive through the on-chip, three-port power divider discussed in the main text which excites the common mode of the JRM by applying rf-voltage of the same polarity and magnitude to two opposite nodes of the JRM, i.e., the top and bottom nodes, the common mode of the JRM can be equivalently excited by applying rf-voltage of the same magnitude but opposite polarity to two adjacent nodes of the JRM. In order to demonstrate this method, we inject the pump drive through the on-chip flux line which in the vicinity of the JRM and at the frequency range relevant to the pump drive (i.e., $16$-$17$ $\operatorname{GHz}$), functions to first order as a short-circuited coupled microstrip line, which capacitively couples to two adjacent nodes of the JRM (in this case 3 and 2) as illustrated in Fig. \ref{PumpFluxLine}. Due to the short-circuit boundary condition set by the semi-loop at the end of the flux-line and the symmetrical coupling between the two coupled traces and nodes 3 and 2 of the JRM, the rf-voltage associated with the pump drive assumes a node at the short, and couples differentially to the two adjacent nodes 3 and 2, which in turn excites the common mode of the structure. 

In Fig. \ref{GainFL}, we exhibit the results of measurements taken on the signal and idler ports of the device centered around $f_S=6.922$ $\operatorname{GHz}$ and $f_I=9.924$ $\operatorname{GHz}$ respectively, which are quite similar to the ones presented in Fig. \ref{Gain}. The primary difference is that the pump drive in this experiment is injected through the flux line instead of the power divider. As seen in the measurements of plots (a), (b) corresponding to the S and I ports, the device gain increases with the applied pump power, and gains in excess of $20$ dB can be achieved on both ports. In plot (c) we exhibit the extracted dynamical bandwidth of the device versus $G^{-1/2}$, which shows that the device follows the amplitude-gain bandwidth product in the limit of large gains in accordance with theory, while plot (d) displays a signal to noise (SNR) improvement measurement taken on the idler port versus gain. By fitting this measurement to the theoretical expression presented in section II while taking into account the noise temperature of the output line $T_{N}=12\pm1$ $\operatorname{K}$, we get an added input noise photon value of $n_{\mathrm{add}}=1/2$ in good agreement with the data, implying that injecting the pump drive through the flux line does not alter the near-quantum-limit performance of the JPC.

It is worth noting that, in these measurements, the pump power applied to the flux line is only about $4$ dB larger than the power applied to the power divider. This implies that the basic low pass filter element (see Fig.1 (e)) implemented on both traces of the on-chip flux line (designed with a cutoff around $3$ $\operatorname{GHz}$) does not effectively block microwave signals in the range $16$-$17$ $\operatorname{GHz}$ used for the pump drive.

In Fig. \ref{DRFL}, we show a maximum input power measurement taken at a power gain of $23$ dB on resonance at the signal (plot(a)) and idler (plot(b)) ports. Both measurements are obtained while flux biasing the JPC and applying the pump drive through the on-chip flux line. In these measurements, the pump drive and the dc current are input to the device using a bias tee at the mixing chamber. The black dashed vertical line indicates the input power of one photon at the signal or idler frequency per inverse dynamical bandwidth of the device. The red open circle marks the $1$ dB compression point at which the device gain drops by $1$ dB.

The results of Figs. \ref{GainFL}, \ref{DRFL} show that the on-chip flux line can serve a dual-purpose of flux-biasing the device and feeding the pump drive, which can be done by combining both control signals at the input of the flux line using a bias tee. However, the main disadvantage of using commercial bias tees, in addition to their size and the need to thermalize them, is that the inductive port of the bias tee is sometimes resistive on the order of a few ohms which limits the maximum dc-current that can be applied to the device without heating the mixing chamber. One potential solution to this problem going forward is realizing a miniature low-loss bias tee on the JPC chip or on a printed circuit board integrated with the JPC.